\documentclass[envcountsame]{mmnp}
\usepackage[tbtags]{amsmath}
\usepackage{amssymb}
\usepackage{graphicx}
\usepackage[latin1]{inputenc}

\newcommand{\noi}{\noindent}

\newcommand{\be}{\begin{equation}}
\newcommand{\ee}{\end{equation}}

\idline{Vol. 8, No. 2}{32}
\doi{10.1051/mmnp/20138202}

\numberwithin{equation}{section}

\begin{document}

\title{Generalized Elastic Model: Fractional Langevin Description, 
Fluctuation Relation and Linear Response}


\author{A. Taloni \inst{1} \sep A. Chechkin \inst{2, 3}  \sep J. Klafter \inst{4} \thanks{\email {alessandro.taloni@gmail.com}} }

\vspace{0.5cm}

\institute{
\inst{1} CNR-IENI, Via R. Cozzi 53, 20125 Milano, Italy\\
\inst{2} Max-Planck-Institute for Physics of Complex Systems, Noethnitzer Str. 38\\ D-91187 Dresden, Germany\\
\inst{3} Akhiezer Institute for Theoretical Physics, NSC KIPT, Kharkov 61108, Ukraine\\ 
\inst{4} School of Chemistry, Tel Aviv University, Tel Aviv 69978, Israel}


\abstract{The Generalized Elastic Model is a linear stochastic model which accounts for the behaviour of many physical systems in nature,  ranging from polymeric chains to single-file systems. If an external perturbation is exerted \emph{only} on  a single point $\vec{x}^\star$ (\emph{tagged probe}), it propagates throughout the entire system. Within the fractional Langevin equation framework, we study the effect of such a perturbation, in cases of a constant force applied. We report most of the results  arising from our previous analysis and, in the present work, we show that the Fox $H$-functions formalism provides a compact, elegant and  useful tool for the study of the scaling properties of any observable. In particular we show how the generalized Kubo fluctuation relations can be expressed in terms of $H$-functions.}





\titlerunning{GLE: FLE, FDT and LR}

\maketitle

\section{Introduction}
\label{sec:Introduction}

The generalized elastic model (GEM) has been firstly introduced in ~\cite{Taloni-FLE} through the following equation:

\begin{equation}
\frac{\partial}{\partial t}\mathbf{h}\left(\vec{x},t\right)=\int d^dx'\Lambda\left(\vec{x}-\vec{x}'\right)\frac{\partial^z }{\partial\left|\vec{x}'\right|^z }\mathbf{h}(\vec{x}',t)+\boldsymbol\eta\left(\vec{x},t\right),
\label{GEM}
\end{equation}

\noi The general formulation in (\ref{GEM}) is given for the  $D$-dimensional stochastic field $\mathbf{h}$ defined in the $d$-dimensional infinite space $\vec{x}$. The white noise $\eta$ 
satisfies the fluctuation-dissipation (FD) relation, i.e.

\be
\langle\eta_{j}\left(\vec{x},t\right)\eta_{k}\left(\vec{x}',t'\right) \rangle
=  2k_BT\Lambda\left(\vec{x}-\vec{x}'\right)\delta_{j\,k}\delta(t-t')
\label{FDT}
\ee

\noi ($j,k\in[1,D]$), where $\Lambda\left(\vec{r}\right)=1/\left|\vec{r}\right|^\alpha$ corresponds to the hydrodynamic friction kernel whose Fourier transform is

\be
\Lambda\left(\vec{q}\right)
=\frac{(4\pi)^{d/2}}{2^{\alpha}}\frac{\Gamma\left((d-\alpha)/2\right)}{\Gamma\left(\alpha/2\right)}\left|\vec{q}\right|^{\alpha-d}=A\left|\vec{q}\right|^{\alpha-d}
\label{hydro_FF}
\ee

\noi if $\frac{d-1}{2}<\alpha<d$.
The fractional derivative  $\partial^z/\partial\left|\vec{x}\right|^z$, defined via its Fourier transform by ~\cite{Zazlawsky}

\be
{\cal F}_{\vec{q}}\left\{\frac{\partial^z}{\partial\left|\vec{x}\right|^z}\right\}\equiv-\left|\vec{q}\right|^z
\label{fractional_operator},
\ee

\noi has another common definition given in term of the Laplacian $\Delta$ as $\frac{\partial^z}{\partial\left|\vec{x}\right|^z}:=-\left(-\Delta\right)^{z/2}$ ~\cite{Samko}. The GEM (\ref{GEM}) accounts for the dynamics of polymers ~\cite{Doi,Granek,semiflexible}, elastic chains \cite{Edwards,Rouse,Zimm},
membranes ~\cite{Granek,membranes,Granek-epl,Zilman,Zilman2} and
rough surfaces ~\cite{interfaces,Krug,surfaces,Majaniemi}, among others. It also reproduces the anomalous diffusive behavior of systems
such as crack propagation ~\cite{Gao} and contact line of a liquid
meniscus ~\cite{DeGennes}. Each one of the above-mentioned physical systems corresponds to a given set of the parameters defining the GEM (\ref{GEM}), namely $z,$ $\alpha$ and $d$, with  $\alpha=d$ in the case of $\Lambda\left(\vec{r}\right)=\delta\left(\left|\vec{r}\right|\right)$ ($A=const$ in (\ref{hydro_FF})).

In Ref.~\cite{Taloni-FLE} we derived a Langevin equation for the probe at a generic position $\vec{x}$ starting from the system Eq.(\ref{GEM}): this equation turned out to be a fractional Langevin equation (FLE), i.e. an usual overdamped Langevin equation  with the normal time derivative replaced by a fractional one.
Generally speaking, in the last decade the number of papers devoted to FLE and its connection to anomalous diffusing systems has notably increased. In ~\cite{Mainardi-Basset} FLE was used to furnish the adequate representation of the Brownian motion of a massive particle moving in a surrounding fluid,  whose  hydrodynamical effects come into play through the Basset-Boussinesq retarding force. Furthermore, in  ~\cite{Lutz} FLE was connected to the fractional Brownian motion  (FBM) ~\cite{Mandelbroot}. 
So far, the most important application of FLE is in protein dynamics. Indeed FLE has been introduced to capture the equilibrium conformational fluctuations of a  protein molecule ~\cite{Kou}. The dynamics of the distance between the donor (\emph{D}) and the acceptor (\emph{A}) coordinates within a protein was shown to be reproduced, with an excellent degree  of accuracy, by a FLE with an applied hookean force. The FLE has been also the subject of theoretical works, which investigated the correlation functions behaviours in presence of an external harmonic   
force ~\cite{Burov,Sau-Fa-PRE}, and in the absence of any ~\cite{Sau-Fa-EPL}. See also the recent review ~\cite{GoychukACP} for more references and details.

Within the framework offered by  GEM (\ref{GEM}), FLE formalism has been successfully applied to the statistical field theory. The stochastic motion of a tagged point (hereafter named \emph{probe} or \emph{tracer} without distinction) on a membrane surface by instance,  was shown to be represented by a FLE ~\cite{Granek-epl}, in the situation on which such a probe was subject to a  linear force mimicking the action of an optical or magnetic tweezer. The tracer height $h\left(\vec{x},t\right)$ is a stochastic field indeed: it is a function of the time $t$ and of the bidimensional membrane parametrization coordinate $\vec{x}$. In single file systems, where a tagged Brownian particle (the probe) undergoes subdiffusive motion on the score of the hard-core collisions with the other identical one dimensional file particles, the FLE was first  phenomenologically proposed ~\cite{Taloni-Lomholt} and afterwards rigorously derived to be the effective tracer stochastic equation ~\cite{Lizana}. Here the probe position is represented by $h\left(x,t\right)$ where $x$ stands for the  particle ordering number along the file. Furthermore, the FLE description of a tagged monomer dynamics in polymeric chains has been recently introduced in ~\cite{Panja}: in this case $\mathbf{h}\left(x,t\right)$ represents the 3 dimensional position of the tagged monomer (probe), and  $x$ its position along the polymer backbone (\emph{curvilinear abscissa}).

In ~\cite{Taloni-PRE} we furnished the analytic expression of the field correlation functions using  the Fox $H$-function formalism \cite{Fox,Mathai,Hilfer}.  
We highlighted the valuable property that physical observables, such as, e.g., mean squared displacement or the structure 
factor, could have a compact and elegant expression in terms of the Fox $H$-functions. These functions gain more and more popularity among 
the scientific community, for their very general nature, which allows to tackle different phenomena in a unified and elegant framework. 
Applications include non-Debye relaxation processes ~\cite{Nonnenmacher}, anomalous diffusion ~\cite{Metzler,Mainardi,Denisov}, reaction-diffusion equations
~\cite{Saxena_2}, relaxation and reaction processes in disordered systems ~\cite{Vlad}, fractional Schroedinger equation ~\cite{Laskin}, to name a few. The book ~\cite{Kilbas} serves a deep analysis of properties of the
Fox $H$-functions. 
Recent monograph ~\cite{Saxena2010} lists many useful properties of the Fox $H$-functions, together with some applications, e.g., in astrophysics. The handbook
~\cite{PrudnikovIII} contains the list of useful properties and integrals of the Fox $H$-functions.  

  In a recent publication ~\cite{Taloni-PRE_2} we added a localized external force to the 
system (\ref{GEM}), namely a force acting \emph{only} on the probe in $\vec{x}^\star$, hereafter called the tagged probe. Under such condition, Eq.(\ref{GEM}) transforms to the following stochastic evolution equation

\begin{equation}
\frac{\partial}{\partial t}\mathbf{h}\left(\vec{x},t\right)=\int d^dx'\Lambda\left(\vec{x}-\vec{x}'\right)
\left[\frac{\partial^z }{\partial\left|\vec{x}'\right|^z }\mathbf{h}(\vec{x}',t)+\mathbf{F}\left\{\mathbf{h}(\vec{x}',t),t\right\}\delta(\vec{x}'-\vec{x}^{\star })\right]+\boldsymbol\eta\left(\vec{x},t\right).
\label{GEM_potential}
\end{equation}

\noi Here $\mathbf{F}\left\{\mathbf{h}(\vec{x},t),t\right\}$ is a force functional of the stochastic fields $\mathbf{h}(\vec{x},t)$ and of the time $t$: it represents the external perturbation applied to the probe particle placed at the position  $\vec{x}^{\star }$. We derived  the FLE for the probes at positions  $\vec{x}^{\star }$ and  $\vec{x}$ (untagged probe).  Our interest is motivated by the increasingly performing nano- and micro-manipulation techniques, that nowadays allow experimentalists to detect the (nano) microscopic fluctuations on a system where a localized external force is applied, as in the case of optical or magnetic tweezer acting on a beads attached to a membrane surface or to a single spot along the polymer backbone ~\cite{Bourdieu,ripple}. We show that the Fox $H$-function formalism constitutes an excellent tool for tackling the problem of the compact representation and the systematic analysis of the scaling properties of any correlation function and drift  when a localized potential is applied to the GEM system (\ref{GEM}). Moreover we demonstrate the usefulness of the Fox $H$-functions in recovering the Einstein and Kubo fluctuation relations.

This article is organized according to the following structure. In Section \ref{sec:Recall} we recall the results formerly obtained  within the FLE framework in the absence of any applied  force. Moreover, we derive the velocity-velocity and position-velocity correlation function in terms of the Fox $H$-functions. In Section \ref{sec:FLE_potential}, starting from
the expression (\ref{GEM_potential}), we derive the FLE equation both for  the tracer
particle placed at position $\vec{x}^{\star }$ and for a particle at a generic position $\vec{x}$, we also study the scaling properties of the noise- and force-propagators.  In  Section
\ref{sec:constant_force} we deal with the situation where the force applied to the probe particle in  $\vec{x}^{\star }$ is constant: we study the scaling properties of the probe average drifts, show the validity of the Kubo fluctuation relations (KFR)  and derive exact results for the Edwards-Wilkinson chain. 
In Appendix \ref{app:FF_prop} we list the Fox $H$-functions properties that we use in our analysis.

%
%
\section{Fractional Langevin Equation framework}
\label{sec:Recall}

In Ref.~\cite{Taloni-FLE} we showed that the statistical properties of the  stochastic systems governed by the equation (\ref{GEM}) can be obtained within the framework of the following fractional Langevin equation  for the probe`s coordinate $h=h_j$ placed at position $\vec{x}$,

\be
K^+\ _{-\infty}D_t^{\beta}h\left(\vec{x},t\right)=\zeta\left(\vec{x},t\right)
\label{FLE}, 
\ee

\noi where 

\be
\beta=\frac{2(z-d)}{\gamma} ,
\label{beta}
\ee

\be
\gamma=2(z+\alpha-d) ,
\label{gamma}
\ee

\noi and 

\be
K^+=\pi^{d/2-1}\frac{\Gamma(d/2)\sin\left(\pi\beta\right)(z+\alpha-d)}{2^{1-d}A^\beta}.
\label{K+}
\ee

\noi The pseudo-differential operator

\be
_aD_t^{\beta}\phi(t)=\frac{1}{\Gamma\left(1-\beta\right)}\frac{d}{dt}\int_{a}^tdt'\frac{1}{\left(t-t'\right)^{\beta}}\phi\left(t'\right),
\     \ 0<\beta<1, \label{RL} 
\ee

\noi represents the left side Riemann-Liouville derivative with lower bound $a<t$ ~\cite{Samko,Podlubny}. We remind that the derivative (\ref{RL}) is equivalent to the Caputo ~\cite{Caputo-der} derivative, i.e.

\be
\frac{d^{\beta}}{dt^\beta}X(t)=\frac{1}{\Gamma(1-\beta)}\int_{a}^{t}dt'\frac{\dot{X}(t')}{(t-t')^\beta}\     \ 0<\beta<1,
\label{Caputo_usual}
\ee

\noi  if  the  lower bound in (\ref{RL}) is set $a=-\infty$ ~\cite{Podlubny}.

 The tracer dynamical representation provided by the equation (\ref{FLE}) is only valid whenever $z>d$, i.e. $\beta<1$ ($(d-1)/2<\alpha<d$). In the case of local hydrodynamics ($\Lambda\left(\vec{r}\right)=\delta\left(\left|\vec{r}\right|\right)$) it is sufficient to set $\alpha=d$ and $A=const$.

\noi As shown in Ref.~\cite{Taloni-PRE},  in spite of their apparent difference, equations (\ref{GEM}) and  (\ref{FLE}) provide the same level of accuracy in the description of the system dynamics. Indeed, while the \emph{Markovian} stochastic behavior of the entire system is ruled by (\ref{GEM}),   Eq.(\ref{FLE}) reproduces the tracer \emph{non-Markovian} fractional Brownian motion. Nonetheless, the overall Markovian and the local non-Markovian representations coincide in furnishing the analytical expression of any system physical observable. This is evident by the properties of the fractional Gaussian noise (fGn) $\zeta\left(\vec{x},t\right)$ appearing on the right hand side of (\ref{FLE}), which is correlated in time \emph{and} in space ~\cite{Taloni-PRE},

\be
\langle \zeta\left(\vec{x},t\right)
\zeta\left(\vec{x}',t'\right)\rangle=\frac{k_BT
K^{+2}A^{2\beta-1}}{2^{\alpha+d}\pi^{(d+1)/2}}\frac{\left|\vec{x}-\vec{x}'\right|^{\alpha}}{\left|t-t'\right|}H_{3\,3}^{2\,2}\left[C\left|{\begin{array}{ccc}
\left(\frac{1}{2},\frac{1}{2}\right)&\left(\beta,\frac{1}{2}\right)& \left(0,\frac{1}{2}\right)\\
\left(-\frac{\alpha}{2},\frac{\gamma}{4}\right)&\left(\beta,\frac{1}{2}\right)&\left(\frac{2-\alpha-d}{2},\frac{\gamma}{4}\right)\\
\end{array} } \right.\right],
\label{FGN_cc}
\ee

\noi where we have introduced the Fox $H$-function  ~\cite{Fox,Mathai,Hilfer}, and defined

\be
C=\frac{2}{A\left|t-t'\right|}\left(\frac{\left|\vec{x}-\vec{x}'\right|}{2}\right)^{\frac{\gamma}{2}}.
\label{C}
\ee

\noi Setting $\vec{x}\to\vec{x}'$ in Eq.(\ref{FGN_cc}) gives the  fluctuation-dissipation relation, i.e. 

\be
\langle \zeta\left(\vec{x},t\right)
\zeta\left(\vec{x},t'\right)\rangle=k_BT \frac{K^+}{\Gamma\left(1-\beta\right)\left|t-t'\right|^{\beta}}.
\label{FLE-FDT}
\ee

\noi Moreover, thanks to Eqs.(\ref{FLE}, \ref{FGN_cc}), the two-point two-time correlation function of the stochastic field $h\left(\vec{x},t\right)$ is expressed as ~\cite{Taloni-PRE}

\be
\begin{array}{l}
\langle h\left(\vec{x},t\right)
h\left(\vec{x}',t'\right)\rangle=\frac{2k_BT
A^{2\beta-1}}{2^{\alpha+d}\gamma\pi^{d/2+1}}\left|\vec{x}-\vec{x}'\right|^{\alpha}\int_{0}^{+\infty}d\omega
\times\\
\       \
\frac{\cos\left(\omega\left(t-t'\right)\right)}{\omega^{2\beta}}
H_{1\,3}^{2\,1}\left[\frac{k}{2}\left|{\begin{array}{ccc}
&\left(\beta,\frac{1}{\gamma}\right)& \\
\left(\frac{-\alpha}{2},\frac{1}{2}\right)&\left(\beta,\frac{1}{\gamma}\right)&\left(\frac{2-\alpha-d}{2},\frac{1}{2}\right)\\
\end{array} } \right.\right].
\end{array}
\label{h_corr_FLE_2}
\ee

\noi  where

\be
k=\left(\frac{|\omega|}{A}\right)^{2/\gamma}\left|\vec{x}-\vec{x}'\right|.
\label{kappa}
\ee

Now, we want to derive the velocity correlation functions, as arising from the FLE (\ref{FLE}). Let us first introduce the  Fourier transform of a function $\mathbf{\phi}\left(\vec{x},t\right)$ as

\be
\mathbf{\phi}\left(\vec{q},\omega\right)=\int_{-\infty}^{+\infty}d^dx \int_{-\infty}^{+\infty}dt\,
\mathbf{\phi}\left(\vec{x},t\right)\,e^{-i\left(\vec{q}\cdot\vec{x}-\omega t\right)}
\label{FFT}
\ee

\noi and introduce the short notation for the time Fourier transform of the force:

\be
{\cal F}_{\omega}\left\{\mathbf{F}\left\{\mathbf{f}(\vec{x}^\star,t),t\right\}\right\}\equiv\int_{-\infty}^{+\infty}dt\,\mathbf{F}\left\{\mathbf{f}(\vec{x}^\star,t),t\right\}
\,e^{i\omega t}
\label{fourier-force}.
\ee

\noi Hence, it is immediate to prove that in the Fourier space

\be
\langle v\left(\vec{x},\omega\right)
v\left(\vec{x}',\omega'\right)\rangle=\frac{1}{K^{+2}}\frac{\langle \zeta\left(\vec{x},\omega\right)
\zeta\left(\vec{x}',\omega'\right)\rangle}{(-i\omega)^{\beta-1}(-i\omega')^{\beta-1}}
\label{app:velocity_VCF_FF_1}
\ee

\noi Thanks to the Fourier transform of the noise correlation function (\ref{FGN_cc}) we can express the former definition 
as

\be
\begin{array}{l}
\langle v\left(\vec{x},\omega\right)
v\left(\vec{x}',\omega'\right)\rangle=\frac{k_BT
A^{2\beta-1}}{2^{\alpha+d-2}\gamma\pi^{d/2-1}}\left|\vec{x}-\vec{x}'\right|^{\alpha}|\omega|^{2-2\beta}\times\\
\          \
\delta(\omega+\omega')H_{1\,3}^{2\,1}\left[\frac{k}{2}\left|{\begin{array}{ccc}
&\left(\beta,\frac{1}{\gamma}\right)& \\
\left(-\frac{\alpha}{2},\frac{1}{2}\right)&\left(\beta,\frac{1}{\gamma}\right)&\left(\frac{2-\alpha-d}{2},\frac{1}{2}\right)\\
\end{array} } \right.\right]
\end{array}.
\label{app:velocity_VCF_FF_2}
\ee

\noi where $k$ has been defined in (\ref{kappa}).

We note that the same expression can be furnished starting from the GEM (\ref{GEM}). Indeed, the solution of (\ref{GEM}) in the Fourier space for the unperturbed systems is

\be
h\left(\vec{q},\omega\right)=\frac{\eta\left(\vec{q},\omega\right)}{-i\omega+A\left|\vec{q}\right|^{\gamma/2}},
\label{app:KFR_sol_FF_unperturbed}
\ee

\noi and for the velocity 

\be
v\left(\vec{q},\omega\right)=\frac{-i\omega\,\eta\left(\vec{q},\omega\right)}{-i\omega+A\left|\vec{q}\right|^{\gamma/2}}.
\label{app:KFR_vel_sol_FF_unperturbed}
\ee

\noi Thanks to  the noise properties in Fourier-Fourier space $\langle \eta\left(\vec{q},\omega\right)\eta\left(\vec{q}',\omega'\right)\rangle=2 k_BT\Lambda\left(\left|\vec{q}\right|\right)(2\pi)^{d+1}\delta(\omega+\omega')\delta(\vec{q}+\vec{q}')$
 we obtain 

\be
\begin{array}{l}
\langle v\left(\vec{x},\omega\right)
v\left(\vec{x}',\omega'\right)\rangle=\frac{4\pi k_BT\delta(\omega+\omega')A\left|\vec{x}-\vec{x}'\right|^{1-d/2}}{(2\pi)^{d/2}}\times\\
\         \
\int_{0}^{+\infty}d\left|\vec{q}\right|\left|\vec{q}\right|^{\alpha-d/2}J_{d/2-1}\left(\left|\vec{q}\right|\left|\vec{x}-\vec{x}^\star\right|\right)\frac{\omega^2}{\omega^2+A^2\left|\vec{q}\right|^{\gamma}},
\label{VCF_FF_int}
\end{array}
\ee

\noi where in addition, we made use of the definition of the $d$-dimensional inverse Fourier transform of an isotropic function $\phi(\left|\vec{q}\right|)$ ~\cite{Champeney}

\be
\begin{array}{l}
\int_{-\infty}^{+\infty}\frac{d^dq}{(2\pi)^d} e^{i\vec{q}\cdot\vec{r}}\phi(\left| \vec{q}\right|)=
\frac{\left|\vec{r}\right|^{1-d/2}}{(2\pi)^{d/2}}\int_{0}^{+\infty}d\left| \vec{q}\right| \left| \vec{q}\right|^{d/2}J_{d/2-1}(\left| \vec{q}\right|\left| \vec{r}\right|)\phi(\left| \vec{q}\right|),
\label{champa}
\end{array}
\ee

\noi with  $J_{d/2-1}$  the Bessel function of  fractional order $d/2-1$. Hence, applying the change of variable  $y=A^{2/\gamma}\left|\vec{q}\right|$  and  recalling that ~\cite{Taloni-PRE},

\be
\frac{1}{1+y^\delta}=\frac{1}{\delta}H_{1\,1}^{1\,1}\left[y\left|{\begin{array}{c}
\left(1,\frac{1}{\delta}\right)\\
\left(1,\frac{1}{\delta}\right)\\
\end{array} } \right.\right],
\label{Fox_frac}
\ee

\noi after employing the property (\ref{pr.5})  we get the expression (\ref{app:velocity_VCF_FF_2}).

Now we want to invert Eq. (\ref{app:velocity_VCF_FF_2}) in the time domain, i.e.

\be
\begin{array}{l}
\langle v\left(\vec{x},t\right)
v\left(\vec{x}',t'\right)\rangle=\frac{k_BT
A^{2\beta-1}}{2^{\alpha+d-1}\gamma\pi^{d/2+1}}\left|\vec{x}-\vec{x}'\right|^{\alpha}\int_0^{\infty}d\omega\,\times\\
\               \
\\cos\left(\omega|t-t'|\right)\omega^{2-2\beta}H_{1\,3}^{2\,1}\left[\frac{k}{2}\left|{\begin{array}{ccc}
&\left(\beta,\frac{1}{\gamma}\right)& \\
\left(-\frac{\alpha}{2},\frac{1}{2}\right)&\left(\beta,\frac{1}{\gamma}\right)&\left(\frac{2-\alpha-d}{2},\frac{1}{2}\right)\\
\end{array} } \right.\right].
\end{array}
\label{app:velocity_VCF_1}
\ee

\noi Note that the integral here does not present any divergence in the limit $\omega\to 0$ as instead of Eq.(\ref{h_corr_FLE_2}). This can be checked by expanding the Fox $H$-function for small argument $k$ thanks to (\ref{short_exp})

\be
H_{1\,3}^{2\,1}\left[\frac{k}{2}\left|{\begin{array}{ccc}
&\left(\beta,\frac{1}{\gamma}\right)& \\
\left(-\frac{\alpha}{2},\frac{1}{2}\right)&\left(\beta,\frac{1}{\gamma}\right)&\left(\frac{2-\alpha-d}{2},\frac{1}{2}\right)\\
\end{array} } \right.\right]\simeq
\frac{2\pi}{\sin(\alpha\pi/\gamma)\Gamma(d/2)}k^{-\alpha}.
\label{H_expansion}
\ee

\noi Inserting the previous expansion in (\ref{app:velocity_VCF_1}) one clearly sees that the power inside the integral is $\propto\omega^{2\alpha/\gamma}$, so that we can proceed to evaluate the expression (\ref{app:velocity_VCF_1}). Making use of (\ref{pr.4}) we arrive at the  expression:

\be
\langle v\left(\vec{x},t\right)
v\left(\vec{x}',t'\right)\rangle=\frac{k_BT
A^{2\beta-1}}{2^{\alpha+d-2+2\beta}\pi^{(d+1)/2}}\frac{\left|\vec{x}-\vec{x}'\right|^{\alpha}}{\left|t-t'\right|^{3-2\beta}}H_{3\,3}^{2\,2}\left[C\left|{\begin{array}{ccc}
\left(-\frac{1}{2}+\beta,\frac{1}{2}\right)&\left(\beta,\frac{1}{2}\right)& \left(-1+\beta,\frac{1}{2}\right)\\
\left(-\frac{\alpha}{2},\frac{\gamma}{4}\right)&\left(\beta,\frac{1}{2}\right)&\left(\frac{2-\alpha-d}{2},\frac{\gamma}{4}\right)\\
\end{array} } \right.\right],
\label{app:velocity_v_corr_FLE_reprise}
\ee

\noi where $C$ has been defined in (\ref{C}). Manipulating the previous form through (\ref{1.2.4}) and (\ref{1.2.5}) one arrives at

\be
\langle v\left(\vec{x},t\right)
v\left(\vec{x}',t'\right)\rangle=\frac{k_BT
A}{2^{d-\alpha}\pi^{(d+1)/2}}\frac{\left|\vec{x}-\vec{x}'\right|^{-\alpha}}{\left|t-t'\right|}H_{3\,3}^{2\,2}\left[C\left|{\begin{array}{ccc}
\left(\frac{1}{2},\frac{1}{2}\right)&\left(1,\frac{1}{2}\right)& \left(0,\frac{1}{2}\right)\\
\left(\frac{\alpha}{2},\frac{\gamma}{4}\right)&\left(1,\frac{1}{2}\right)&\left(1+\frac{\alpha-d}{2},\frac{\gamma}{4}\right)\\
\end{array} } \right.\right],
\label{v_corr_FLE}
\ee

\noi If we want to calculate the autocorrelation function we must set $\vec{x}\to\vec{x}'$. Expanding for small $k$  the Fox $H$-function in (\ref{v_corr_FLE}) by means of (\ref{short_exp}), we arrive at the expression  

\be
\langle v\left(\vec{x},t\right)
v\left(\vec{x},t'\right)\rangle=-\frac{\alpha k_BT
A^{\beta}2^{3-d}}{\gamma^2\pi^{d/2}}\frac{\Gamma\left(1-\beta\right)}{\Gamma\left(d/2\right)\left|t-t'\right|^{2-\beta}},
\label{v_autocorr_FLE}
\ee

\noi  from which it is found that the  field velocity  is always anti-correlated, as it is expected for subdiffusive fractional Brownian systems ~\cite{Mandelbroot}.

The next step that we want to take is the calculation of the position-velocity correlation function $\langle h\left(\vec{x},t\right)v\left(\vec{x}',t'\right)\rangle$. We do it within the framework of GEM (\ref{GEM}).
  From (\ref{app:KFR_sol_FF_unperturbed}), (\ref{app:KFR_vel_sol_FF_unperturbed}) and (\ref{champa}) we achieve

\be
\langle h\left(\vec{x},t\right)v\left(\vec{x}',t'\right)\rangle=\frac{2k_BT\,A\left|\vec{x}-\vec{x}^\star\right|^{1-d/2}}{(2\pi)^{d/2}}\int_{0}^{+\infty}d\left|\vec{q}\right|\left|\vec{q}\right|^{\alpha-d/2}J_{d/2-1}\left(\left|\vec{q}\right|\left|\vec{x}-\vec{x}'\right|\right)\int_{-\infty}^{+\infty}\frac{d\omega}{2\pi} \frac{i\omega\,e^{-i\omega \left|t-t'\right|}}{A^2\left|\vec{q}\right|^\gamma+\omega^2}
\label{app:KFR_h-v_cc_2}
\ee

\noi which, inverted in time, gives

\be
\langle h\left(\vec{x},t\right)v\left(\vec{x}',t'\right)\rangle=\frac{k_BT\,A\left|\vec{x}-\vec{x}^\star\right|^{1-d/2}}{(2\pi)^{d/2}}\int_{0}^{+\infty}d\left|\vec{q}\right|\left|\vec{q}\right|^{\alpha-d/2}J_{d/2-1}\left(\left|\vec{q}\right|\left|\vec{x}-\vec{x}'\right|\right)e^{-A\left|\vec{q}\right|^{\gamma/2}\left|t-t'\right|}
\label{app:KFR_h-v_cc_3}
\ee

\noi Recalling that the exponential function has an  $H$-representation which is

\be
e^{-y}=H_{0\,1}^{1\,0}\left[y\left|{\begin{array}{c}
- \\
\left(0,1\right)\\
\end{array} } \right.\right],
\label{exp_FF}
\ee

\noi  and using the property (\ref{pr.5}), the Fox $H$-function expression for the above correlation function is 

\be
\langle h\left(\vec{x},t\right)v\left(\vec{x}^\star,t'\right)\rangle=\frac{k_BTA\left|\vec{x}-\vec{x}^\star\right|^{-\alpha}}{2^{d-\alpha}\pi^{d/2}}H_{2\,1}^{1\,1}\left[A\left|t-t'\right|\left(\frac{2}{\left|\vec{x}-\vec{x}^\star\right|}\right)^{\gamma/2}\left|{\begin{array}{ccc}
\left(1-\frac{\alpha}{2},\frac{\gamma}{4}\right) & \left(\frac{d-\alpha}{2},\frac{\gamma}{4}\right)\\
\left(0,1\right) & (0,1)\\
\end{array} } \right.\right].
\label{FLEx_constant_h-v_cc_Fox}
\ee

%
%
\section{Fractional Langevin Equation with applied force}
\label{sec:FLE_potential}

For consistency, we hereby remind briefly the derivation of the FLE for the probe particle
placed both at the position $\vec{x}^\star$ (tagged probe), where the external force is applied, and at the generic position $\vec{x}$ (untaged probe) ~\cite{Taloni-PRE_2}.

\noi In the Fourier space the solution of (\ref{GEM_potential}) is obtained as

\be
\mathbf{h}\left(\vec{q},\omega\right)=\frac{A\,{\cal F}_{\omega}\left\{\mathbf{F}\left\{\mathbf{h}(\vec{x}^\star,t),t\right\}\right\}e^{-i\vec{q}\cdot\vec{x}^\star}}{\left|\vec{q}\right|^{d-\alpha}\left(-i\omega+A\left|\vec{q}\right|^{\gamma/2}\right)}+\frac{\boldsymbol{\eta}\left(\vec{q},\omega\right)}{-i\omega+A\left|\vec{q}\right|^{\gamma/2}},
\label{sol_FF}
\ee

\noi where we  made use of the definition (\ref{hydro_FF}). 
\noi We then multiply both sides of (\ref{sol_FF}) by
$K^+(-i\omega)^{\beta}$, where $K^+$ and $\beta$ have been defined respectively in (\ref{K+}) and (\ref{beta}):

\begin{equation}
\begin{array}{l}
K^+(-i\omega)^{\beta}\mathbf{h}\left(\vec{q},\omega\right)={\cal F}_{\omega}\left\{\mathbf{F}\left\{\mathbf{h}(\vec{x}^\star,t),t\right\}\right\}\frac{AK^+(-i\omega)^{\beta} e^{-i\vec{q}\cdot\vec{x}^\star}}{\left|\vec{q}\right|^{d-\alpha}\left(-i\omega+A\left|\vec{q}\right|^{\gamma/2}\right)}
+\frac{K^+(-i\omega)^{\beta}}{-i\omega+A\left|\vec{q}\right|^{\gamma/2}}\,\boldsymbol{\eta}\left(\vec{q},\omega\right).
\label{sol_FF_1}
\end{array}
\end{equation}

\noi We  now derive the FLE for the tracer  at a generic position $\vec{x}$: to proceed further we first have to invert in the space domain both terms on the right side of (\ref{sol_FF_1}). For the first  we get

\begin{equation}
\begin{array}{l}
{\cal F}_{\omega}\left\{\mathbf{F}\left\{\mathbf{h}(\vec{x}^\star,t),t\right\}\right\}AK^+(-i\omega)^{\beta}\int_{-\infty}^{+\infty}\frac{d^dq}{(2\pi)^d}\frac{ e^{i\vec{q}\cdot\left(\vec{x}-\vec{x}^\star\right)}}{\left|\vec{q}\right|^{d-\alpha}\left(-i\omega+A\left|\vec{q}\right|^{\gamma/2}\right)}=\\
{\cal F}_{\omega}\left\{\mathbf{F}\left\{\mathbf{h}(\vec{x}^\star,t),t\right\}\right\}\frac{AK^+(-i\omega)^{\beta}\left|\vec{x}-\vec{x}^\star\right|^{1-d/2}}{(2\pi)^{d/2}}
\int_{0}^{+\infty}d\left|\vec{q}\right|\frac{\left|\vec{q}\right|^{\alpha-d/2}J_{d/2-1}\left(\left|\vec{q}\right|\left|\vec{x}-\vec{x}^\star\right|\right) }{-i\omega+A\left|\vec{q}\right|^{\gamma/2}}  .
\label{sol_FF_2}
\end{array}
\end{equation}

\noi Defining the following function as 

\be
\Theta\left(\left|\vec{x}\right|,\omega\right)=
\frac{AK^+(-i\omega)^{\beta}\left|\vec{x}\right|^{1-d/2}}{(2\pi)^{d/2}}\int_{0}^{+\infty}d\left|\vec{q}\right|\frac{\left|\vec{q}\right|^{\alpha-d/2}J_{d/2-1}\left(\left|\vec{q}\right|\left|\vec{x}\right|\right) }{-i\omega+A\left|\vec{q}\right|^{\gamma/2}}  ,
\label{Theta}
\ee

\noi we obtain that the first term in the right hand side of Eq.(\ref{sol_FF_1}) after the inverse Fourier transform in space is given by

\be
{\cal F}_{\omega}\left\{\mathbf{F}\left\{\mathbf{h}(\vec{x}^\star,t),t\right\}\right\}\Theta\left(\left|\vec{x}-\vec{x}^\star\right|,\omega\right)
\label{sol_FF_1term}.
\ee

\noi The second term can be treated in the same way  ~\cite{Taloni-FLE,Taloni-PRE}. Indeed, inverting in the space domain we find 

\be
\boldsymbol{\zeta}\left(\vec{x},\omega\right)=\int_{-\infty}^{+\infty}d\vec{x}'\boldsymbol{\eta}\left(\vec{x}',\omega\right)\Phi\left(\left|\vec{x}'-\vec{x}\right|,\omega\right)
\label{sol_FF_2term}
\ee

\noi where, according to (\ref{champa}), the function $\Phi\left(\left|\vec{x}\right|,\omega\right)$ is defined as

\be
\Phi\left(\left|\vec{x}\right|,\omega\right)=
\frac{K^+(-i\omega)^{\beta}\left|\vec{x}\right|^{1-d/2}}{(2\pi)^{d/2}}\int_{0}^{+\infty}d\left|\vec{q}\right|\frac{\left|\vec{q}\right|^{d/2}J_{d/2-1}\left(\left|\vec{q}\right|\left|\vec{x}\right|\right) }{-i\omega+A\left|\vec{q}\right|^{\gamma/2}}.
\label{Phi}
\ee

\noi By combining Eqs.(\ref{sol_FF_1term}) and (\ref{sol_FF_2term}) we can write Eq.(\ref{sol_FF_1})
after its invertion in the space domain as

\be
K^+(-i\omega)^{\beta}\mathbf{h}\left(\vec{x},\omega\right)=
{\cal F}_{\omega}\left\{\mathbf{F}\left\{\mathbf{h}(\vec{x}^\star,t),t\right\}\right\}\Theta\left(\left|\vec{x}-\vec{x}^\star\right|,\omega\right) + \boldsymbol{\zeta}\left(\vec{x},\omega\right)
\label{FLE_x_omega}
\ee
\noi It is convenient to use Eq.(\ref{FLE_x_omega}) as a starting point to calculate the response
for the untagged and tagged probes in the subsequent Section.

The inverse Fourier transforms of  $\Theta\left(\left|\vec{x}\right|,\omega\right)$ and
$\Phi\left(\left|\vec{x}\right|,\omega\right)$ are performed  as follows.  Let us start from the force propagator $\Theta\left(\left|\vec{x}\right|,t\right)$. The key point here is the representation of 
the Riemann-Liouville fractional derivative (\ref{RL}) in Fourier space ~\cite{Podlubny},

\be
\int_{-\infty}^{+\infty}dt \,e^{i\omega t}\,_{-\infty}D_t^{\beta}\phi(t)=(-i\omega)^{\beta}\phi(\omega).
\label{app:Theta_RL_FT}
\ee

\noi Hence we invert in time the expression (\ref{Theta}) as

\be
\Theta\left(\left|\vec{x}\right|,t\right)=\frac{AK^+\left|\vec{x}\right|^{1-d/2}}{(2\pi)^{d/2}}\int_{0}^{+\infty}d\left|\vec{q}\right|\left|\vec{q}\right|^{\alpha-d/2}J_{d/2-1}\left(\left|\vec{q}\right|\left|\vec{x}\right|\right)_{-\infty}D_t^{\beta}\left(e^{-A\left|\vec{q}\right|^{\gamma/2}t}\theta(t)\right)
\label{app:Theta_Theta}.
\ee

\noi where $\theta(t)$ represents the Heaviside step function.

\noi From Eq.(\ref{Phi}) it results that the noise-propagator $\Phi\left(\left|\vec{x}\right|,t\right)$ has the same structure as $\Theta\left(\left|\vec{x}\right|,t\right)$. We can then treat it in the same way, obtaining

\be
\Phi\left(\left|\vec{x}\right|,t\right)=\frac{AK^+\left|\vec{x}\right|^{1-d/2}}{(2\pi)^{d/2}}\int_{0}^{+\infty}d\left|\vec{q}\right|\left|\vec{q}\right|^{d/2}J_{d/2-1}\left(\left|\vec{q}\right|\left|\vec{x}\right|\right)_{-\infty}D_t^{\beta}\left(e^{-A\left|\vec{q}\right|^{\gamma/2}t}\theta(t)\right)
\label{app:Theta_Phi}.
\ee

\noi Taking into account the definition (\ref{RL}) we rewrite Eqs.(\ref{app:Theta_Theta}) and (\ref{app:Theta_Phi}) as 

\be
\Theta\left(\left|\vec{x}\right|,t\right)=
\frac{AK^+\left|\vec{x}\right|^{1-d/2}}{(2\pi)^{d/2}}   \,_0D_t^\beta\int_{0}^{+\infty}d\left|\vec{q}\right|\left|\vec{q}\right|^{\alpha-d/2}J_{d/2-1}\left(\left|\vec{q}\right|\left|\vec{x}\right|\right) e^{-A\left|\vec{q}\right|^{\gamma/2}t}   ,
\label{Theta_time}
\ee

\noi and

\be
\Phi\left(\left|\vec{x}\right|,t\right)=
\frac{AK^+\left|\vec{x}\right|^{1-d/2}}{(2\pi)^{d/2}} 
\,_0D_t^\beta\int_{0}^{+\infty}d\left|\vec{q}\right|\left|\vec{q}\right|^{d/2}J_{d/2-1}\left(\left|\vec{q}\right|\left|\vec{x}\right|\right) e^{-A\left|\vec{q}\right|^{\gamma/2}t}   ,
\label{Phi_time}
\ee

\noi where $_0D_t^\beta$ represents the Riemann-Liouville fractional derivative (\ref{RL}) with the lower bound $a=0$. In Eqs.(\ref{Theta_time}) and (\ref{Phi_time}) $t$
is non-negative,
otherwise Eqs.(\ref{app:Theta_Theta}) and (\ref{app:Theta_Phi}) give zero. Furthermore we  note that the two integral expressions (\ref{Theta_time}) and (\ref{Phi_time}) coincide in the special case of local 
hydrodynamic interactions. 

The two propagators can be expressed in terms of the Fox $H$-functions by the use of Eq.(\ref{exp_FF}),
the property (\ref{pr.5}) and the fractional derivative of the Fox $H$-function (\ref{frac_der})
~\cite{Nonnenmacher}:

\be
\Theta\left(\left|\vec{x}\right|,t\right)=\frac{A^{1+\beta}K^+2^{\gamma/2-d}}{\pi^{d/2}\left|\vec{x}\right|^{\gamma/2}}H_{3\,2}^{1\,2}\left[2^{\gamma/2}\frac{t}{\tau\left(\left|\vec{x}\right|\right)}\left|{\begin{array}{ccc}
\left(-\beta,1\right) & \left(1-\frac{\gamma}{4},\frac{\gamma}{4}\right) & \left(\frac{2d-z-\alpha}{2},\frac{\gamma}{4}\right)\\
\left(-\beta,1\right) & (0,1)\\
\end{array} } \right.\right]
\label{Theta_time_Fox}.
\ee

\noi and

\be
\Phi\left(\left|\vec{x}\right|,t\right)=\frac{A^{1+\beta}K^+2^{z-d}}{\pi^{d/2}\left|\vec{x}\right|^{z}}H_{3\,2}^{1\,2}\left[2^{\gamma/2}\frac{t}{\tau\left(\left|\vec{x}\right|\right)}\left|{\begin{array}{ccc}
      \left(-\beta,1\right) & \left(1-\frac{z}{2},\frac{\gamma}{4}\right) & \left(\frac{d-z}{2},\frac{\gamma}{4}\right)\\
\left(-\beta,1\right) & (0,1)\\
\end{array} } \right.\right]
\label{Phi_time_Fox}
\ee

\noi where  the \emph{correlation time} has been introduced as ~\cite{Taloni-PRE_2,Taloni-EPL}

\be
 \tau\left(\left|\vec{x}\right|\right)=\frac{\left|\vec{x}\right|^{\gamma/2}}{A}.
\label{tau}
\ee

\noi Finally, applying the inverse Fourier transformation in time to Eq. (\ref{FLE_x_omega}), we get the form of the FLE for the particle at a generic position $\vec{x}$:

\be
K^+\,_{-\infty}D_t^{\beta}\mathbf{h}\left(\vec{x},t\right)=
\int_{-\infty}^{t}dt'\mathbf{F}\left\{\mathbf{h}(\vec{x}^\star,t'),t'\right\}\Theta\left(\left|\vec{x}-\vec{x}^\star\right|,t-t'\right)+\boldsymbol{\zeta}\left(\vec{x},t\right) ,
\label{FLE_x} 
\ee

\noi where the non-Markovian noise

\be
\boldsymbol{\zeta}\left(\vec{x},t\right)=\int_{-\infty}^{+\infty}d\vec{x}'\int_{-\infty}^{t}dt'\boldsymbol{\eta}\left(\vec{x}',t'\right)\Phi\left(\left|\vec{x}'-\vec{x}\right|,t-t'\right)
\label{sol_2term}
\ee

\noi fulfills fluctuation-dissipation relation (\ref{FLE-FDT}). The analysis of (\ref{FLE_x}) shows that the
probe particle at the position $\vec{x}$ undergoes an effective force that, compared to the force acting on $\vec{x}^\star$, is shifted in time  and in space according to the function defined in (\ref{Theta_time_Fox}). Indeed,  $\Theta\left(\left|\vec{x}-\vec{x}^\star\right|,t-t'\right)$ can be seen as the propagator carrying the external perturbation exerted at the point $\vec{x}^\star$ at time $t'$,  to the point $\vec{x}$ at time $t$. Likewise, the function $\Phi\left(\left|\vec{x}-\vec{x}^\star\right|,t-t'\right)$ represents the propagator of the Brownian random source $\boldsymbol{\eta}\left(\vec{x},t\right)$ from the point $\vec{x}'$ to the point $\vec{x}$ in the time elapsed between $t'$ and $t$: the sum of the contributions arising from the whole system within the interval $t'\in[-\infty,t]$ generates stationary non-Markovian fGn $\boldsymbol{\zeta}\left(\vec{x},t\right)$.

We point out that in  Eqs.(\ref{FLE_x}) and (\ref{sol_2term})  time integrals extend to $t$ instead of $+\infty$ as formally required by the definition of the inverse Fourier transform. Indeed, both propagators $\Theta\left(\left|\vec{x}\right|,t\right)$ and $\Phi\left(\left|\vec{x}\right|,t\right)$ differ from zero only for $t>0$, as it was already mentioned. Moreover, equations (\ref{Theta_time_Fox}) and (\ref{Phi_time_Fox}) highlight the valuable property for which the propagators attain different regimes whether the time $t$ is larger or smaller than the  characteristic correlation time $\tau$. As a matter of fact, the Fox $H$-function formalism provides a comprehensive mathematical formulation of a function showing double scaling behavior, as any Fox $H$-function exhibits different regimes for small and large values of its argument ~\cite{Saxena2010,PrudnikovIII}.     
The time $\tau\left(\left|\vec{x}\right|\right)$ can be interpreted as the time needed for the \emph{information} (either an external perturbation or the same random source $\boldsymbol{\eta}$) to travel within a distance $\left|\vec{x}\right|$ along the elastic system ~\cite{Taloni-PRE_2,Taloni-EPL}. Similarly, the \emph{correlation length} $\xi(t)= (At)^{2/\gamma}$ represents the distance within which the points are mutually influencing their stochastic dynamics ~\cite{interfaces,Krug,Barabasi, Meakin,Family}. The two scaling regimes exhibited by the propagators (\ref{Theta_time_Fox}) and (\ref{Phi_time_Fox})  have  strong implications in the two-point two-time correlation functions behavior ~\cite{Taloni-EPL} and, most important, in the probes responses to external perturbation, as demonstrated below.

We now turn to the derivation of the FLE for the tagged tracer at $\vec{x}^\star$. In this case it is sufficient to take the limit $\vec{x}\to \vec{x}^\star$ in (\ref{FLE_x}). Taking expression (\ref{Theta}), we recall that the Bessel function expansion for small argument is ~\cite{Abramowitz}

\be
J_{d/2-1}(r)\sim\frac{1}{\Gamma(d/2)}\left(\frac{2}{r}\right)^{1-d/2}
\label{Bessel_exp}
\ee

\noi from which we have

\be
\Theta\left(0,\omega\right)=
\frac{AK^+(-i\omega)^{\beta}}{2^{d-1}(\pi)^{d/2}\Gamma(d/2)}\int_{0}^{+\infty}d\left|\vec{q}\right|\frac{\left|\vec{q}\right|^{\alpha-1}}{-i\omega + A\left|\vec{q}\right|^{\gamma/2}}
\label{Theta_0}.
\ee

\noi Changing variable $y=\left|\vec{q}\right|^{\gamma/2}$, solving the 
integral according to  Ref.~\cite{PrudnikovI}, and  thanks to the definition of $K^+$ (\ref{K+}) we obtain that

\be
\Theta\left(0,\omega\right)=1
\label{Theta_0_2}
\ee

\noi  which, substituted in (\ref{sol_FF_1term}) and inverted in time, gives the  FLE expression  for the probe particle placed at $\vec{x}^\star$ : 

\be
K^+D_C^{\beta}\mathbf{h}\left(\vec{x}^\star,t\right)=\mathbf{F}\left\{\mathbf{h}(\vec{x}^\star,t),t\right\}+\boldsymbol{\zeta}\left(\vec{x}^\star,t\right).
\label{FLE_xstar} 
\ee

%
%
\section{Constant force: $F_0\,\theta(t)$}
\label{sec:constant_force}

The situation that we are going to address in this paper concerns an external constant force applied to the tagged probe $\vec{x}^\star$, i.e.

\be
\mathbf{F}\left\{\mathbf{h}(\vec{x}^\star,t),t\right\}=F_0\theta(t).
\label{force_constant}
\ee

\noi where $F_0$ represents the force along one direction (say $F_0\equiv F_j$) and $\theta(t)$ is the Heaviside step function. We are interested in the average drift of the tagged and untagged tracers in $\vec{x}$ and $\vec{x}^\star$, namely $\langle h\left(\vec{x},t\right)\rangle_{F_0}$ and  $\langle h\left(\vec{x}^\star,t\right)\rangle_{F_0}$ respectively. We first focus on the untagged probe. The starting point is Eq.(\ref{FLE_x_omega}), where the Fourier transform of the force is given by ~\cite{Gelfand}

\be
{\cal F}_{\omega} \left\{F_0\theta(t)\right\}= F_0\left\{\frac{i}{\omega} + \pi\delta(\omega)\right\}
\label{gelfand},
\ee

\noi while $\Theta$ and $\boldsymbol{\zeta}$ are given by Eqs.(\ref{Theta}) and (\ref{sol_FF_2term}), respectively. Using the fact that $\delta(\omega)\Theta\left(\left|\vec{x}\right|,\omega\right)$ = 0,
we get the following equation for the Fourier transform of the average drift in time,

\be
-i\omega\langle h\left(\vec{x},\omega\right)\rangle_{F_0}=
\frac{A\left|\vec{x}\right|^{1-d/2}}{(2\pi)^{d/2}}F_0\int_{0}^{+\infty}d\left|\vec{q}\right|\frac{\left|\vec{q}\right|^{\alpha-d/2}J_{d/2-1}\left(\left|\vec{q}\right|\left|\vec{x}\right|\right) }{-i\omega+A\left|\vec{q}\right|^{\gamma/2}}
\label{drift_omega}
\ee

\noi Taking an inverse Fourier transform in time we get

\be
\frac{d\langle h\left(\vec{x},t\right)\rangle_{F_0}}{dt}=
\frac{A\left|\vec{x}-\vec{x}^\star\right|^{1-d/2}}{(2\pi)^{d/2}}F_0
\int_{0}^{+\infty}d\left|\vec{q}\right|\left|\vec{q}\right|^{\alpha-d/2}J_{d/2-1}\left(\left|\vec{q}\right|\left|\vec{x}-\vec{x}^\star\right|\right)e^{-A\left|\vec{q}\right|^{\gamma/2} t}\theta(t).
\label{drift_time}
\ee

\noi Integrating both sides of Eq.(\ref{drift_time}) from $0$ to $t$ and noting that  
$\langle h\left(\vec{x},t=0\right)\rangle_{F_0}=0$, we arrive at the following expression for the 
average drift,

\be
\langle h\left(\vec{x},t\right)\rangle_{F_0}=\frac{A\left|\vec{x}-\vec{x}^\star\right|^{1-d/2}}{(2\pi)^{d/2}}F_0\int_0^t dt'
\int_{0}^{+\infty}d\left|\vec{q}\right|\left|\vec{q}\right|^{\alpha-d/2}J_{d/2-1}\left(\left|\vec{q}\right|\left|\vec{x}-\vec{x}^\star\right|\right)e^{-A\left|\vec{q}\right|^{\gamma/2} t'}.
\label{FLEx_constant_force_average_drift}
\ee

\noi  First we cast the exponential according to (\ref{exp_FF}),

\be
\begin{array}{l}
\langle h\left(\vec{x},t\right)\rangle_{F_0}=\\
\frac{A\left|\vec{x}-\vec{x}^\star\right|^{1-d/2}}{(2\pi)^{d/2}}F_0\int_0^t dt'\int_{0}^{+\infty}d\left|\vec{q}\right|\left|\vec{q}\right|^{\alpha-d/2}J_{d/2-1}\left(\left|\vec{q}\right|\left|\vec{x}-\vec{x}^\star\right|\right) H_{0\,1}^{1\,0}\left[A\left|\vec{q}\right|^{\alpha+z-d}t'\left|{\begin{array}{ccc}
- \\
\left(0,1\right)\\
\end{array} } \right.\right]
\label{FLEx_constant_force_average_drift_Fox_1}.
\end{array}
\ee

\noi Thanks to the integral (\ref{pr.5}) it gets the form 

\be
\langle h\left(\vec{x},t\right)\rangle_{F_0}=\frac{A\left|\vec{x}-\vec{x}^\star\right|^{-\alpha}}{2^{d-\alpha}\pi^{d/2}}F_0\int_0^tdt'\,H_{2\,1}^{1\,1}\left[2^{\gamma/2}\frac{t}{\tau\left(\left|\vec{x}-\vec{x}^\star\right|\right)}\left|{\begin{array}{ccc}
\left(1-\frac{\alpha}{2},\frac{\gamma}{4}\right) & \left(\frac{d-\alpha}{2},\frac{\gamma}{4}\right)\\
\left(0,1\right) \\
\end{array} } \right.\right]
\label{FLEx_constant_force_average_drift_Fox}.
\ee

\noi Finally, after taking the integral over $t'$ using (\ref{1.16.4.1}), we get

\be
\langle h\left(\vec{x},t\right)\rangle_{F_0}=\frac{\left|\vec{x}-\vec{x}^\star\right|^{z-d}}{2^{z}\pi^{d/2}}F_0\,H_{3\,2}^{1\,2}\left[2^{\gamma/2}\frac{t}{\tau\left(\left|\vec{x}-\vec{x}^\star\right|\right)}\left|{\begin{array}{ccc}
\left(1,1)(1+\frac{z-d}{2},\frac{\gamma}{4}\right) & \left(\frac{z}{2},\frac{\gamma}{4}\right)\\
\left(1,1)(0,1\right) \\
\end{array} } \right.\right]
\label{FLEx_constant_force_average_drift_Fox_final}.
\ee

\noi Taking here the limit $\vec{x}\to \vec{x}^\star$ and using the expansion (\ref{long_exp}) 
 we arrive at the expression for the tagged probe drift,

\be
\langle h\left(\vec{x}^\star,t\right)\rangle_{F_0}\simeq\frac{2^{1-d/2}A^{\beta}\Gamma(1-\beta)}{(2\pi)^{d/2}\Gamma(d/2)(z-d)}F_0\,t^{\beta}  ,
\label{FLEx_constant_force_drift_tagged_probe}
\ee

\noi where $\beta$ is given by Eq.(\ref{beta}). The motion of the  untagged particle is more interesting, indeed its  average drift attains
 two differrent regimes whether  $t\ll\tau$ or $t\gg\tau$ respectively, in correspondence with the short and long time behavior of the Fox function appearing in Eq.(\ref{FLEx_constant_force_average_drift_Fox_final}). In the upcoming subsection we study the behaviour for short and long times of systems presenting long-range and local hydrodynamic interactions respectively.

\subsection{$\langle h\left(\vec{x},t\right)\rangle_{F_0}$ scaling behavior}
\label{F_const_scaling}

Firstly, our analysis will concern with long range hydrodynamic systems.

\begin{itemize}
\item $\mathbf{t\ll} \boldsymbol\tau$. We start by using the Fox $H$-function expression (\ref{FLEx_constant_force_average_drift_Fox_final}): by expanding the Fox $H$-function according to (\ref{short_exp}) we obtain that for time shorter than correlation time  $\tau$ the drift is given by

\be
\langle h\left(\vec{x},t\right)\rangle_{F_0}\simeq\frac{A\left|\vec{x}-\vec{x}^\star\right|^{-\alpha}}{2^{d-\alpha}\pi^{d/2}}\frac{\Gamma(\alpha/2)}{\Gamma\left(\frac{d-\alpha}{2}\right)}F_0\,t
\label{FLEx_constant_force_average_drift_hydro_short}.
\ee

\item $\mathbf{t\gg} \boldsymbol\tau$. The long time limit of the untagged probe average can be obtained by expanding (\ref{FLEx_constant_force_average_drift_Fox_final}) with the use of (\ref{long_exp}):

\be
\langle h\left(\vec{x},t\right)\rangle_{F_0}\simeq\langle h\left(\vec{x}^\star,t\right)\rangle_{F_0}
\simeq\frac{2^{1-d/2}A^{\beta}\Gamma(1-\beta)}{(2\pi)^{d/2}\Gamma(d/2)(z-d)}F_0\,t^{\beta}
\label{FLEx_constant_force_average_drift_hydro_long}.
\ee

\end{itemize}

\noi Now we consider local  hydrodynamic interactions, hence we set  $A=const$, $\alpha=d$, $\gamma=2z$ and $\beta=\frac{z-d}{z}$ in (\ref{FLEx_constant_force_average_drift_Fox_final}). 

\begin{itemize} 

\item $\mathbf{t\ll}\boldsymbol \tau$. An expansion of (\ref{FLEx_constant_force_average_drift_Fox_final}) according to (\ref{short_exp})  gives, when $z\neq 2m$  
($m\in \mathbb{N}$),

\be
\langle h\left(\vec{x},t\right)\rangle_{F_0}\simeq\frac{2^{z-2}\Gamma(\frac{z+d}{2})\Gamma(\frac{z}{2})}{\pi^{d/2+1}}\frac{z\sin\left(\frac{z\pi}{2}\right)}{\left|\vec{x}-\vec{x}^\star\right|^{z+d}} F_0(At)^2,
\label{FLEx_constant_force_average_drift_local_short}.
\ee

\noi Equation(\ref{FLEx_constant_force_average_drift_local_short}) points out that
for $2+4m<z<4+4m$ the response of the probe is opposite to the
external disturbance $F_0$, while for  $4m<z<2+4m$ they have the
same sign  ~\cite{Taloni-PRE_2}. When $z=2m$ the response is slower than any power so that
we expect ~\cite{Majaniemi, Taloni-EPL}

\be
\langle h\left(\vec{x},t\right)\rangle_{F_0}\propto F_0
\frac{t^{\beta+1}}{\left|\vec{x}-\vec{x}^\star\right|^{z}}e^{-\frac{\left|\vec{x}-\vec{x}^\star\right|^{1/\beta}}{(At)^{1/(z-d)}}}
\label{FLEx_constant_force_average_drift_local_short_marg}.
\ee

\item $\mathbf{t\gg}\boldsymbol \tau$.  For long times we can expand the expression (\ref{FLEx_constant_force_average_drift_Fox_final}) and get the same results as for long-range
hydrodynamic system, see Eq.(\ref{FLEx_constant_force_average_drift_hydro_long}).

\end{itemize}

\noi We can summarize the results obtained in this section in the following compact form

\be
\langle h\left(x,t\right)\rangle_{F_0} =F_0\frac{\left|\vec{x}-\vec{x}^\star\right|^{z-d}}{(2\pi)^{d/2}}f\left[\frac{t}{\tau\left(\left|\vec{x}-\vec{x}^\star\right|\right)}\right]
\label{scaling_average_drift}
\ee

\noi The scaling function $f\left[u\right]$ exhibits two distinct behaviours whether $u\ll 1$ or $u\gg 1$. From (\ref{FLEx_constant_force_average_drift_hydro_short}), (\ref{FLEx_constant_force_average_drift_local_short}) and (\ref{FLEx_constant_force_average_drift_local_short_marg}), it turns out that when $u\ll 1$
 
\be
f\left[u\right]\left\{
\begin{array}{ccc}
\sim2^{\alpha-d/2}\frac{\Gamma\left(\frac{\alpha}{2}\right)}{\Gamma\left(\frac{d-\alpha}{2}\right)}u &   & i)\\
\sim\frac{2^{z+d/2-2}}{\pi}z\sin\left(\frac{z\pi}{2}\right)\Gamma\left(\frac{z}{2}\right)\Gamma\left(\frac{z+d}{2}\right)u^2 &   & ii)\\
\propto u^{\beta+1}e^{-u^{1/(d-z)}} &   & iii)
\end{array}\right.
\label{f_sht_th}
\ee

\noi for $i)$ long range, $ii)$ local ($z\neq 2m$) and $iii)$ local ($z= 2m$) hydrodynamic interactions, respectively. When $u\gg 1$ we have invariably 

\be
f\left[u\right]\simeq\frac{2^{1-d/2}}{z-d}\frac{\Gamma(1-\beta)}{\Gamma\left(\frac{d}{2}\right)}u^\beta
\label{f_lot_th}.
\ee

\subsection{Kubo fluctuation relations}
\label{F_const_kubo}

In this subsection we show how the previous results satisfy the Kubo fluctuation relations ~\cite{Taloni-PRE_2,Kubo-book,Villamaina} which connect the average drift in presence of an external disturbance $F_0$, to the correlation function in absence of any. Firstly we introduce the two-point two-time correlation function as

\be
\begin{array}{l}
\langle \delta_{t} h\left(\vec{x},t\right)
\delta_{t'} h\left(\vec{x}',t'\right)\rangle=\frac{k_BT}{(2\pi)^{d/2}}\left|\vec{x}-\vec{x}'\right|^{z-d}\times\\
\left\{f\left[\frac{t}{\tau\left(\left|\vec{x}-\vec{x}'\right|\right)}\right]+f\left[\frac{t'}{\tau\left(\left|\vec{x}-\vec{x}'\right|\right)}\right]-f\left[\frac{\left|t-t'\right|}{\tau\left(\left|\vec{x}-\vec{x}'\right|\right)}\right]\right\}
\label{2P_2T_CF_th},
\end{array}
\ee

\noi where $\delta_{t} h\left(\vec{x},t\right)=h\left(\vec{x},t\right)-h\left(\vec{x},0\right)$ and $f[u]$ has been introduced in (\ref{scaling_average_drift}) ~\cite{Taloni-EPL}. From (\ref{scaling_average_drift}) and (\ref{2P_2T_CF_th}) it follows that the average drift is given by

\be
\langle h\left(x,t\right)\rangle_{F_0} =\frac{F_0}{2k_BT}\langle \delta_{t} h\left(\vec{x},t\right)
\delta_{t} h\left(\vec{x}^\star,t\right)\rangle.
\label{GKFR}
\ee

\noi The former equality constitutes the most general form of the Kubo fluctuation relation (KFR) and encompasses both tagged and untagged particles. Another formulation of the KFR is furnished as follows. Comparing the expression for the untagged probe drift (\ref{FLEx_constant_force_average_drift_Fox}) with the corresponding expression of the position-velocity correlation function (\ref{FLEx_constant_h-v_cc_Fox}), we recover 
the generalized KFR such as ~\cite{Taloni-PRE_2,Kubo-book,Villamaina}

\be
\langle h\left(\vec{x},t\right)\rangle_{F_0}=\frac{F_0}{k_BT}\int_0^tdt'\langle h\left(\vec{x},t\right)v\left(\vec{x}^\star,0\right)\rangle
\label{KFR}
\ee

\noi which corresponds to (\ref{GKFR}) after integration. In case of the tagged probe we get the (usual)
Einstein relation ~\cite{Taloni-PRE_2}

\be
\langle h\left(\vec{x}^\star,t\right)\rangle_{F_0}=\frac{ \langle\delta_t^2h\left(\vec{x}^\star,t\right)\rangle}{2k_BT}F_0
\label{GER}
\ee

\noi which corresponds to the limit $\vec{x}\to\vec{x}^\star$ of (\ref{GKFR}). One can see that the Einstein relation (\ref{GER}) is fulfilled also by untagged probe, provided that $t\gg \tau$, i.e. by the time that the correlation length $\xi(t)= (At)^{2/\gamma}$ exceeds the distance  $\left|\vec{x}-\vec{x}^\star\right|$. 

To sum up, Eq.(\ref{GKFR})  constitutes the Kubo fluctuation relation for the linear system (\ref{GEM})  in the case of a  local constant force 
applied in $\vec{x}^\star$. The FLE (\ref{FLE}) can thus be considered as a stochastic representation of the  the KFR   (\ref{GKFR}) with the random force  $\zeta(\vec{x},t)$ satisfying the  (second) FD relation (\ref{FLE-FDT}).

\subsection{Edward-Wilkinson chain}

\noi We now test some of the previous results in case of a system with local hydrodynamic interactions and $z=2$,  $d=1$. In this special case  Eq.(\ref{GEM_potential}) reads

\begin{equation}
\frac{\partial}{\partial t}h\left(x,t\right)=A\frac{\partial^2 }{\partial x^2 }h(\vec{x},t)+F_0\theta(t)\delta(x^\star-x)+\eta\left(\vec{x},t\right),
\label{EW}
\end{equation}

\noi where $\beta=1/2$, $\gamma=4$, $A=const>0$ and $K^+=2/\sqrt{A}$ from (\ref{beta}), (\ref{gamma}) and (\ref{K+}), respectively. This is, for instance, the equation for the 1-dimensional Rouse polymer ~\cite{Rouse} or the Edward-Wilkinson chain ~\cite{Edwards} and it constitutes a benchmark for our analysis since the drift
can be calculated directly from Eq.(\ref{EW}) in terms of simpler functions.

We start by considering the tagged probe, Eq.(\ref{FLEx_constant_force_drift_tagged_probe}) yields

\be
\langle h\left(\vec{x}^\star,t\right)\rangle_{F_0}=F_0\sqrt{\frac{At}{\pi}}.
\label{EWxstar_constant_force_average_drift}
\ee

\noi For the untagged probe, we make use of Eq.(\ref{FLEx_constant_force_average_drift_Fox}) plugging in  the values of parameters specified above. By using the reduction formula (\ref{8.3.2.6}), the formulae (\ref{pr.1}) and (\ref{1.125}), we obtain

\be
H_{2\,1}^{1\,1}\left[\frac{t}{\xi}\left(\frac{2}{\left|\vec{x}-\vec{x}^\star\right|}\right)^{2}\left|{\begin{array}{ccc}
\left(\frac{1}{2},1\right) & \left(0,1\right)\\
\left(0,1\right)\\
\end{array} } \right.\right]=
\frac{\left|\vec{x}-\vec{x}^\star\right|}{2}\sqrt{\frac{1}{At}}e^{-\frac{\left(\vec{x}-\vec{x}^\star\right)^2}{4At}}.
\label{FF_EW}
\ee

\noi After the integration in time  we get the final expression

\be
\langle h\left(\vec{x},t\right)\rangle_{F_0}=F_0\left[\sqrt{\frac{At}{\pi}}e^{-\frac{\left|\vec{x}-\vec{x}^\star\right|^2}{4At}}-\frac{\left|\vec{x}-\vec{x}^\star\right|}{2}\,erfc\left(\frac{\left|\vec{x}-\vec{x}^\star\right|}{2\sqrt{At}}\right)\right],
\label{EWx_constant_force_average_drift}
\ee

\noi where $erfc$ represents the complementary error function. The same result can be obtained
directly from Eq.(\ref{EW}) without using Fox $H$-function formalism. We can then study short and long 
time limits of the former equation. For $t\ll\frac{\left(\vec{x}-\vec{x}^\star\right)^2}{A}$ we have

\be
\begin{array}{l}
\langle h\left(\vec{x},t\right)\rangle_{F_0}\simeq \frac{2}{\sqrt{\pi}\left|\vec{x}-\vec{x}^\star\right|^2}F_0\left(At\right)^{3/2}e^{-\frac{\left|\vec{x}-\vec{x}^\star\right|^2}{4At}},
\label{EWx_constant_force_average_drift_short}
\end{array}
\ee

\noi which matches the expression (\ref{FLEx_constant_force_average_drift_local_short_marg}). 
For $t\gg\frac{\left|\vec{x}-\vec{x}^\star\right|^2}{A}$ instead the asymptotic drift is given by
Eq.(\ref{EWxstar_constant_force_average_drift}).

%
%
\section{Conclusions}
\label{conclusions}

In this paper we derived the FLE for tagged and untagged probe particle in a generalized elastic model where a localized force is supposed to operate. The tagged probe is considered to be the point on which the external force acts, while the untagged probe is any other point on the system, which is secondarily affected by the action of the perturbation. Within the FLE framework, the stochastic motion of both tracers evidently appears to be influenced  
by the external force, since the external perturbation propagates through the system. This propagation is mathematically expressed by  the noise and force propagators (Green's functions), carrying the perturbation between two points of the system, in a certain lapse of time. We analyzed the double scaling behaviour of these propagators which arise naturally thanks to Fox $H$-function formalism.  We have shown that such
behaviour affects the stochastic dynamics of both tracers. In particular, we demonstrated how the response of the untagged tracer differs drastically for long and short time, in the case of a constant force applied, when hydrodynamic interactions can be considered long-ranged or local.

\noi We related our general theoretical set-up to the specific case of Edward-Wilkinson chain with a local constant force. Importantly, we have shown how the Fox $H$-function formalism constitutes a powerful, compact 
and elegant way to express the observables and allows the straightforward study of their asymptotic behaviours.

\appendix{Appendix}
%
%
\section{Fox function properties}
\label{app:FF_prop}

In this Section we enumerate the properties of the Fox $H$-functions
that we use throughout our analysis. This list is  not an exhaustive compendium of the Fox $H$-function properties, for which the reader could
refer to ~\cite{Mathai,Hilfer,PrudnikovIII}. Some useful properties are also  reported in ~\cite{Taloni-PRE}.

\noi Fox $H$-functions are defined through the Mellin transform 

\be
H_{p\,q}^{m\,n}\left[y\left|{\begin{array}{ccc}
(a_1,A_1)&...&(a_p,A_p)\\
(b_1,B_1)&...&(b_q,B_q)\\
\end{array} } \right.\right]=\frac{1}{2\pi i}\int_L\chi(s)y^{-s}ds
\label{FF_def}
\ee

\noi with $1\leq m\leq q$, $0\leq n\leq p$.  $\chi(s)$ is given by

\be
\chi(s)=\frac{\prod_{j=1}^{m}\Gamma\left(b_j+B_js\right)\prod_{j=1}^{n}\Gamma\left(1-a_j-A_js\right)}{\prod_{j=m+1}^{q}\Gamma\left(1-b_j-B_js\right)\prod_{j=n+1}^{p}\Gamma\left(a_j+A_js\right)}.
\label{Mellin_def}
\ee

\noi where $A_j$ and $B_j$ are positive numbers while  $a_j$ and
$b_j$ are complex. Empty products are interpreted as being unity.

\noi For convenience in this Section we adopt the following short notation

\be
H_{p\,q}^{m\,n}\left[y\left|{\begin{array}{ccc}
(a_1,A_1)&...&(a_p,A_p)\\
(b_1,B_1)&...&(b_q,B_q)\\
\end{array} } \right.\right]=H_{p\,q}^{m\,n}\left[y\left|{\begin{array}{c}
\left[a_p,A_p\right]\\
\left[b_q,B_q\right]\\
\end{array} } \right.\right].
\label{FF_short}
\ee

The useful rules are hereafter listed:

\be
H_{p\,q}^{m\,n}\left[ y\left|{\begin{array}{c}
\left[a_p,A_p\right]\\
\left[b_q,B_q\right]\\
\end{array} } \right.\right]=H_{q\,p}^{n\,m}\left[ \frac{1}{y}\left|{\begin{array}{c}
\left[1-b_q,B_q\right]\\
\left[1-a_p,A_p\right]\\
\end{array} } \right.\right]  ,
\label{pr.1}
\ee

\be
H_{p\,q}^{m\,n}\left[ y\left|{\begin{array}{c}
\left[a_p,A_p\right]\\
\left[b_q,B_q\right](a_1,A_1)\\
\end{array} } \right.\right]=H_{p-1\,q-1}^{m\,n-1}\left[ y\left|{\begin{array}{c}
(a_2,A_2),....,(a_p,A_p)\\
\left[b_{q-1},B_{q-1}\right]\\
\end{array} } \right.\right]   ,
\label{8.3.2.6}
\ee

\be
\begin{array}{c}
\int_0^{\infty}dy\,y^{\alpha-1}\cos(\sigma y)H_{p\,q}^{m\,n}\left[\omega y^r\left|{\begin{array}{c}
\left[a_p,A_p\right]\\
\left[b_q,B_q\right]\\
\end{array} } \right.\right]=\\
\frac{2^{\alpha-1}\sqrt{\pi}}{\sigma^{\alpha}}H_{p+2\,q}^{m\,n+1}\left[\omega \left(\frac{2}{\sigma}\right)^r\left|{\begin{array}{c}
\left(\frac{2-\alpha}{2},\frac{r}{2}\right)\left[a_p,A_p\right]\left(\frac{1-\alpha}{2},\frac{r}{2}\right)\\
\left[b_q,B_q\right]\\
\end{array} } \right.\right]   ,
\end{array}
\label{pr.4}
\ee

\be
\begin{array}{c}
\int_0^{x}dy\,y^{\alpha-1}H_{p\,q}^{m\,n}\left[\omega y\left|{\begin{array}{c}
\left[a_p,A_p\right]\\
\left[b_q,B_q\right]\\
\end{array} } \right.\right]=\\
x^{\alpha}H_{p+1\,q+1}^{m\,n+1}\left[\omega x\left|{\begin{array}{c}
\left(1-\alpha,1\right)\left[a_p,A_p\right]\\
\left[b_q,B_q\right]\left(-\alpha,1\right)\\
\end{array} } \right.\right]  ,
\end{array}
\label{1.16.4.1}
\ee

\be
\frac{1}{k}H_{p\,q}^{m\,n}\left[y\left|{\begin{array}{c}
\left[a_p,A_p\right]\\
\left[b_q,B_q\right]\\
\end{array} } \right.\right]=
H_{p\,q}^{m\,n}\left[y^k\left|{\begin{array}{c}
\left[a_p,kA_p\right]\\
\left[b_q,kB_q\right]\\
\end{array} } \right.\right]  ,
\label{1.2.4}
\ee

\be
y^{\sigma}H_{p\,q}^{m\,n}\left[y\left|{\begin{array}{c}
\left[a_p,A_p\right]\\
\left[b_q,B_q\right]\\
\end{array} } \right.\right]=
H_{p\,q}^{m\,n}\left[y\left|{\begin{array}{c}
\left[a_p+\sigma A_p,A_p\right]\\
\left[b_q+\sigma B_q,B_q\right]\\
\end{array} } \right.\right]  ,
\label{1.2.5}
\ee

\be
\begin{array}{c}
\int_0^{\infty}dy\,y^{\alpha-1}J_{\nu}(\sigma y)H_{p\,q}^{m\,n}\left[\omega y^r\left|{\begin{array}{c}
\left[a_p,A_p\right]\\
\left[b_q,B_q\right]\\
\end{array} } \right.\right]=\\
\frac{2^{\alpha-1}}{\sigma^{\alpha}}H_{p+2\,q}^{m\,n+1}\left[\omega \left(\frac{2}{\sigma}\right)^r\left|{\begin{array}{c}
\left(1-\frac{\alpha+\nu}{2},\frac{r}{2}\right)\left[a_p,A_p\right]\left(1-\frac{\alpha-\nu}{2},\frac{r}{2}\right)\\
\left[b_q,B_q\right]\\
\end{array} } \right.\right]  ,
\end{array}
\label{pr.5}
\ee

\be
\begin{array}{l}
_{0}D_t^\nu y^{\alpha}H_{p\,q}^{m\,n}\left[(\sigma y)^\beta\left|{\begin{array}{c}
\left[a_p,A_p\right]\\
\left[b_q,B_q\right]\\
\end{array} }
  \right.\right]=\\
\      \
\omega^{\alpha-\nu}H_{p+1\,q+1}^{m\,n+1}\left[(\sigma y)^\beta\left|{\begin{array}{cc}
(-\alpha,\beta)&\left[a_p,A_p\right]\\
\left[b_q,B_q\right]&(\nu-\alpha,\beta)\\
\end{array} }
  \right.\right]  ~\cite{Nonnenmacher},
\label{frac_der}
\end{array}
\ee

\be
\begin{array}{c}
H_{0\,1}^{1\,0}\left[y\left|{\begin{array}{c}
-\\
\left[b_q,B_q\right]\\
\end{array} } \right.\right]=\frac{y^{\frac{b}{B}}}{B}e^{-y^{1/B}}  .
\end{array}
\label{1.125}
\ee

\noi Asymptotic expansion: $y\to 0$,

\be
\begin{array}{l}
H_{p\,q}^{m\,n}\left[y\left|{\begin{array}{c}
\left[a_p,A_p\right]\\
\left[b_q,B_q\right]\\
\end{array} }
  \right.\right]=\sum_{i=1}^{m}\sum_{k=0}^{\infty}c_{ik}\frac{(-1)^{k}}{k!B_i}y^{\frac{b_i+k}{B_i}}\\
c_{ik}=\frac{\prod_{j=1,j\neq i}^{m}\Gamma\left(b_j-\frac{(b_i+k)B_j}{B_i}\right)\prod_{j=1}^{n}\Gamma\left(1-a_j+\frac{(b_i+k)A_j}{B_i}\right)}{\prod_{j=m+1}^{q}\Gamma\left(1-b_j+\frac{(b_i+k)B_j}{B_i}\right)\prod_{j=n+1}^{p}\Gamma\left(a_j-\frac{(b_i+k)A_j}{B_i}\right)}  .
\label{short_exp}
\end{array}
\ee 

\noi This expansion is valid whenever
$\sum_{i=1}^qB_i-\sum_{i=1}^pA_i\geq 0$ or $\sum_{i=1}^qB_i-\sum_{i=1}^pA_i< 0$ and $\sum_{i=1}^nA_i-\sum_{i=n+1}^pA_i+\sum_{i=1}^mB_i-\sum_{i=m+1}^qB_i> 0$.

\noi Asymptotic expansion: $y\to \infty$,

\be
\begin{array}{l}
H_{p\,q}^{m\,n}\left[y\left|{\begin{array}{c}
\left[a_p,A_p\right]\\
\left[b_q,B_q\right]\\
\end{array} }
  \right.\right]=\sum_{i=1}^{n}\sum_{k=0}^{\infty}c_{ik}\frac{(-1)^{k}}{k!A_i}y^{\frac{-(1-a_i+k)}{A_i}}\\
c_{ik}=\frac{\prod_{j=1,j\neq i}^{n}\Gamma\left(1-a_j-\frac{(1-a_i+k)A_j}{A_i}\right)\prod_{j=1}^{m}\Gamma\left(b_j+\frac{(1-a_i+k)B_j}{A_i}\right)}{\prod_{j=n+1}^{p}\Gamma\left(a_j+\frac{(1-a_i+k)A_j}{A_i}\right)\prod_{j=m+1}^{q}\Gamma\left(1-b_j-\frac{(1-a_i+k)B_j}{A_i}\right)}  .
\label{long_exp}
\end{array}
\ee 

\noi This expansion is valid whenever
$\sum_{i=1}^qB_i-\sum_{i=1}^pA_i\leq 0$ or $\sum_{i=1}^qB_i-\sum_{i=1}^pA_i> 0$ and $\sum_{i=1}^nA_i-\sum_{i=n+1}^pA_i+\sum_{i=1}^mB_i-\sum_{i=m+1}^qB_i> 0$. Empty products are interpreted as being unity.



\begin{acknowledgement} 
A.T. acknowledges the financial support of the European Complexity-net pilot project ``LOCAT''. 
\end{acknowledgement}



\end{document}